# Electron Loss to the Continuum in the Projectile ionization for Positronium – Helium Atom Collision

Susmita Roy and C. Sinha

Theoretical Physics Department, Indian Association for the Cultivation of Science,

Kolkata – 700 032, India.

The dynamics of the electron loss to the continuum ( ELC ) from the light neutral projectile positronium ( Ps ) atom in collision with the He atom is studied in the framework of the post collisional Coulomb Distorted Eikonal Approximation ( CDEA ). Both the fully differential ( TDCS ) and the double differential ( DDCS ) cross sections are investigated in the intermediate and high incident energies. Results are compared with the existing experiment and other theories , where possible.

## 1. Introduction

Electron emission process in atom – atom or ion - atom collisions becomes particularly interesting at the same time complex when a structured projectile loses electron in collision with the target. Two independent channels can contribute to such projectile electron loss process ( commonly known as ELC ) e.g., the projectile electron can be knocked out by the screened target nucleus or by a target electron [ 1 ]. In the former process ( singly inelastic ) the target usually remains in its ground state i.e., target elastic while in the latter ( doubly inelastic) , the target also gets excited or ionized i.e., target inelastic . Since these two channels lead to different final products, their contributions are to be added incoherently ( i.e., in the cross section level ) . The relative importance of the two channels depends on the incident energy as well as on the particular collision system.

Since the pioneering experimental discovery [ 2 ] of the ELC, a significant number of experimental [ 3 - 14 ] and theoretical studies [ 15 - 25 ] were performed on the projectile electron loss process ( ELC ) in different ion – atom , atom – atom collisions. In all the measurements [ 1 - 14 ] of such process, a prominent cusp shaped ( broad ) peak , depending on the kinematics was observed in the angular ( energy ) distributions of the ejected electron . This peak was attributed to the electron loss from the projectile ion / atom into its low - lying continuum, usually referred to as the ELC peak ( electron loss peak ). Proper theoretical description of the ELC peak in respect of magnitude, position asymmetry etc. is still now a challenge to the theorists.

Until very recently, experimental [ 2 – 10, 12 ] and theoretical [ 15 – 22 ] investigations on the ELC process were mostly limited to bare , partially stripped [ 2 - 6, 15 , 16, 19 - 22 ] or neutral [ 7 - 10, 12, 17, 18 ] heavy projectiles. The first observation on the ELC process by light neutral projectile due to Armitage et al [ 11 ] for the Ps - He atom system stimulated theoretical workers [ 23 – 26 ] to venture the study of this process. The basic difference between the heavy projectile and the light projectile impact ELC phenomena is that, in the former case the deflection as well as the energy loss of the projectile, due to its heavy mass is negligibly small leading to a pronounced peak / cusp in the forward direction , while in the latter case , the light projectile can scatter to large angles and its energy loss is also not negligible leading to a broad ELC peak / cusp. In both the cases, the ELC phenomenon occurs particularly when the ejected electron and the scattered projectile are very close to each other in the velocity space ($v_e \approx v_p$). Study of the dynamics e.g., angular and energy distributions of the ELC process gives valuable information about the ionizing mechanisms and provides a unique insight into the collision dynamics as well into the atomic structure of the collision partners. In the experiment of Armitage et al [ 11 ] , apart from the absolute break up cross sections, the longitudinal positron energy distributions were also measured in arbitrary units ( not absolute ) in search of ELC.

Regarding the theoretical situation for this process , only a limited number of works were reported [ 23 – 26 ] following the experiment [ 11 ] probably because of the complexity lying with the five body system . Sarkadi [ 23 ] studied the ELC phenomena in the framework of the Classical Trajectory Monte Carlo ( CTMC ) model . Later, Starrett et al [ 25 , 26 ] performed a quantal calculation in the frame work of Impulse Approximation ( IA ) restoring to the so called peaking approximation , supposed to be reliable at high incident energies. The qualitative agreement of the quantal [ 25, 26 ] calculations with the existing experiment [ 11 ] was quite good , although a significant quantitative discrepancy was noted in some cases.

The present work addresses the target elastic break up process:

$$( e^+ e ) + He ( 1s ) \rightarrow e^+ + e^- + He ( 1s ) \qquad (I)$$

with the motivation for a detailed study of the decoupled angular and the energy distributions of both the $e^+$ and the $e^-$, the need for which was already emphasized in the experimental work

[ 11 ] . Both the ELC and the asymmetric ionization processes are studied giving particular emphasis on the former one ( ELC ) . Since the electrons of the He atom are much more tightly bound than the electron of the Ps atom, the probability of the electron loss from the projectile Ps is expected to be much higher than the ionization of the target. Due to the large excitation energy of the He atom we have neglected any virtual or real excitation of the He target during the fragmentation.

The Ps – He break up process is essentially a five body problem. The theoretical prescription of such a process is quite difficult since both the initial components of the reaction ( I ) are composite bodies and proper inclusion of the electron exchange effect is therefore even more difficult [ 27, 28 ] . As such , one has to resort to some simplifying assumptions for the theoretical modeling of such a many body ( five body ) reaction process . The present calculation is performed in the frame work of the post collisional Coulomb Distorted Eikonal Approximation ( CDEA ) taking account of the proper asymptotic three body boundary condition in the final channel and the full three body interaction is also incorporated in the final channel which is highly crucial for a proper theoretical description of such ELC process. We also consider the electron exchange effect between the projectile and the target electrons in the frame work of a simplified model similar to the Ochkur Rudge Approximations [ 27 - 29 ] in order to remedy the difficulties of the Born Oppenheimer approximation arising from the non orthogonality of the wave functions .

## 2. Theory

The prior form of the ionization amplitude for the aforesaid process is given as :

$$T_{if}^{prior} = \langle \Psi_f^- | V_i | \psi_i \rangle \quad (1a)$$

$$f = \langle \Psi_f^-(\vec{r}_1, \vec{r}_2, \vec{r}_3, \vec{r}_4) | V_i | \psi_i(\vec{r}_1, \vec{r}_2, \vec{r}_3, \vec{r}_4) \rangle \quad (1b)$$

$$g = \langle \Psi_f^-(\vec{r}_1, \vec{r}_3, \vec{r}_2, \vec{r}_4) | V_i | \psi_i(\vec{r}_1, \vec{r}_2, \vec{r}_3, \vec{r}_4) \rangle \quad (1c)$$

where $f$ and $g$ are the direct and exchange amplitudes respectively. The initial asymptotic state solution $\psi_i$ occurring in equation (1a) is chosen as:

$$\psi_i = \phi_{Ps}(|\vec{r}_1 - \vec{r}_2|) \, e^{(i\vec{k}_i \cdot \vec{\rho})} \phi_{He}(\vec{r}_3, \vec{r}_4) \quad (2)$$

where $\rho = (\vec{r}_1 + \vec{r}_2)/2$; $\vec{r}_1$, $\vec{r}_2$ are the position vectors of the positron and the electron of the Ps and $\vec{r}_3, \vec{r}_4$ are the position vectors of the two bound electrons of the He atom with respect to the target nucleus. $\vec{k}_i$ being the incident momentum of the Ps atom. $\phi_{Ps}$ and $\phi_{He}$ are the respective ground state wave functions of the Ps [ 24 ] and the He [ 30 ] atom. $V_i$ in equation ( 1a ) is the perturbation in the initial channel which is the part of the total interaction not diagonalized in the initial state and is given as:

$$V_i = \frac{Z_t}{r_1} - \frac{Z_t}{r_2} - \frac{1}{r_{13}} + \frac{1}{r_{23}} - \frac{1}{r_{14}} + \frac{1}{r_{24}} \quad (3)$$

where $\vec{r}_{13} = \vec{r}_1 - \vec{r}_3$, $\vec{r}_{23} = \vec{r}_2 - \vec{r}_3$, $\vec{r}_{14} = \vec{r}_1 - \vec{r}_4$, $\vec{r}_{24} = \vec{r}_2 - \vec{r}_4$,

$Z_t$ is the charge of the target nucleus.

$\Psi_f^-$ in equation (1a) is an exact solution of the five body problem satisfying the incoming - wave boundary condition. In the present prescription the full five body final state wave function $\Psi_f^-$ is approximated in the framework of the CDEA by the following ansatz:

$$\Psi_f^-(\vec{r}_1, \vec{r}_2, \vec{r}_3, \vec{r}_4) = (2\pi)^{-3/2} \phi_{He}(\vec{r}_3, \vec{r}_4) e^{i\vec{k}_1 \cdot \vec{r}_1} e^{i\vec{k}_2 \cdot \vec{r}_2} N_{\vec{k}_{12}}$$
$$\,_1F_1[-i\alpha_{12}, 1, -i(k_{12} r_{12} + \vec{k}_{12} \cdot \vec{r}_{12})] \chi_f(\vec{r}_1, \vec{r}_2) \quad (4a)$$

where $\alpha_{12} = -\dfrac{1}{\mu |\vec{k}_{12}|}$, $\vec{k}_{12} = (\vec{k}_1 - \vec{k}_2)/2$,

$$N_{\vec{k}_{12}} = \exp\left(-\dfrac{\pi \alpha_{12}}{2}\right) \Gamma(1 + i\alpha_{12}), \quad (4b)$$

$\mu$ being the electron - positron reduced mass, $\vec{k}_1, \vec{k}_2$ are the final momenta of the $e^+$ and the $e^-$ respectively. In equation (4a) $\chi_f(\vec{r}_1, \vec{r}_2)$ represents the distorted wave function of the outgoing $e^+$ and the $e^-$ in the frame work of the eikonal approximation [24, 31] and assumes the following form: $\chi_f(\vec{r}_1, \vec{r}_2) = (r_1 + r_{1Z})^{-i\eta_1} (r_2 + r_{2Z})^{i\eta_2}$; $\eta_1 = \dfrac{Z_t}{k_1}$

$\eta_2 = \dfrac{Z_t}{k_2}$ and $r_{1Z}, r_{2Z}$ are the z components of the respective vectors $\vec{r}_1$ and $\vec{r}_2$.

In constructing the final state wave function, the following assumptions are made. Since the process studied is the target elastic, the target electrons are considered to be passive in the interaction with the outgoing $e^-$ and the $e^+$ and the distorting potential felt by these two particles is due to the target nucleus only, thereby reducing the five body problem to a three body one in the final channel. This approximation is supposed to be legitimate unless the collision energy reaches a very high value and for a not too heavy atomic target [ 31 ]. However, the influence of the atomic electrons on the projectile ionization is considered through the first order perturbation interaction ( $V_i$ in equations 1a – 1c ).

After much analytical reduction [ 32, 33 ], the ionization amplitude $T_{if}$ in equation ( 1a ) is finally reduced to a two dimensional numerical integral [ 34 ] where $T_{if}$ includes the direct as well as the exchange amplitude.

$$T_{if} = \left| f - g \right|^2 \qquad (5)$$

The exchange effect is considered in the frame work of Ochkur Rudge approximation [ 27 - 29 ].

The triple differential cross section ( TDCS ) for the ionization process is given by :

$$\frac{d^3\sigma}{dE_1 \, d\Omega_1 \, d\Omega_2} = \frac{k_1 \, k_2 \, \mu_i}{k_i} \left| T_{if} \right|^2 \qquad (6)$$

The corresponding expression for the double differential cross section ( DDCS ) is obtained by integrating the TDCS in equation ( 6 ) over the solid angle of the positron ( $\theta_1$, $\phi_1$ ) or the electron ( $\theta_2$, $\phi_2$ ) and is given by :

$$\frac{d^2\sigma}{dE_2 \, d\Omega_{2,1}} = \frac{k_1 \, k_2 \, \mu_i}{k_i} \int \left| T_{if} \right|^2 d\Omega_{1,2} \qquad (7)$$

## 3. Results and Discussions

The triple ( TDCS ) and the double ( DDCS ) differential cross sections are computed for the ELC phenomena at intermediate and high incident energies with respect to the threshold ( 6.8 eV ) of the process ( I ) using the coplanar geometry. For the TDCS curves we adopt the conventional notation [ 24 ]. Present results are compared with the available theory [ 25 ] and experiment [ 11 ] , where possible. It should be mentioned here that the experimental DDCS refers to the longitudinal energy distributions of the $e^+$ ( only ) measured in arbitrary units while the present DDCS represent the conventional DDCS for both the $e^+$ and the $e^-$ .

Figure 1 shows the angular distributions ($\theta_1$) of the scattered positron ( TDCS ) at the incident positronium energy 33 eV for some fixed values of the ejected electron angle $\theta_2$. All the curves exhibit an ELC cusp around $\theta_1 = \theta_2$. The well-known Coulomb density factor [ 35 ] is responsible for such singular structures. The qualitative behaviour of the present TDCS more or less agrees with the theoretical findings [ 25 ], displayed in the inset of the fig. 1 .

Figure 2 elucidates the $e^+$ angular distributions ( TDCS ) at different incident energies for the forward ( $\theta_2 = 0^0$ ) emission of the electron. Apart from the broad ELC cusp around $\theta_1 = 0^0$, the $e^+$ distribution also shows a rise in the backward directions. The Ps breakup cross section is found to increase with increasing incident energy ( fig.2 ), as is expected in an ionization process. The cusp around the forward scattering angle ($\theta_1 = 0^0$) of the $e^+$ becomes broader with decreasing incident energy, since due to its light mass the $e^+$ is preferentially scattered to a higher angle at lower scattering energies.

Figures 3a – 3c demonstrate the energy distributions ( DDCS ) of the $e^-$ and the $e^+$ at different incident energies. In these cases the DDCS refers to the summation over the scattering angles ($\theta_1, \phi_1$) of the positron for a fixed emission angle ($\theta_2 = 0^0$) of the electron barring the fig 3c which also includes the DDCS corresponding to some other higher ejection angles. As is apparent from the figures, the DDCS exhibits a broad peak slightly below ( above ) half of the residual collision energy $E_{res}/2$ in the outgoing $e^-$ ( $e^+$ ) energy spectra . The shifting of the DDCS peak from the $E_{res}/2$ corroborates qualitatively the experimental findings for heavy particle [ 12 ] as well as for Ps impact [ 11, 36 ] ionization process, although in the latter , the shift is in the opposite direction. However, it may be pointed out here that the experimental [ 11 ] data for the process ( I ) are expressed in arbitrary units ( not absolute ).

Fig. 3a also includes the First Born Approximation ( FBA ) DDCS results extracted from the present computer code. It is evident from the figures that the present eikonal DDCS are qualitatively in sharp contrast to the FBA results in respect of the position of the ELC peak. In fact, the present DDCS peak shifts towards the lower ejected energy from $E_{res}/2$ while both in the FBA and IA [ 25 ] the situation is just the reverse. It should be pointed out here that the FBA ( unlike the present model ) does not take account of the final channel distortion due to the target nucleus , responsible for the shift of the DDCS peak [ 12 ] . Further, it is well known that the FBA is not supposed to be adequate at lower incident energies. It is also indicated from the figures ( 3a – 3c ) that the inclusion of electron exchange effect does not modify the qualitative behaviour of the curves.

Figure 3c clearly reflects a strong $e^- - e^+$ asymmetry in the DDCS as was also observed in the experiment [ 11, 36 ]. As may be noted from fig. 3c , the lower electron energy is favoured in the $e^-$ energy spectrum while in the positron energy spectrum , the reverse is true, i.e., the higher positron energy is preferred. Further, this asymmetry dies out with increasing incident energy ( eg., 200 eV, 500 eV in fig 3b ) , corroborating qualitatively the other existing theoretical results [ 25, 26 ]. A plausible physical explanation for the apparent $e^- - e^+$ asymmetry in the energy spectrum ( DDCS ) as well as its behaviour with respect to the incident energy could be given as follows. In the post collisional interaction, the $e^-$ and the $e^+$ are distorted by their increasing interactions with the target. Since the $e^+$ feels repulsion while the $e^-$ feels attraction due to the short range interaction with the target nucleus, on an average the $e^-$ remains closer to the target while the $e^+$ moves away from it. As such, the probability of the $e^-$ ( $e^+$ ) to suffer hard ( soft ) collisions with the target increases with decreasing incident energy. Thus in the post collisional effect , the electron is in the combined field of its parent ( $e^+$ ) and the target nucleus indicating that the description of the ELC process is beyond single center and is the outcome of a two center effect.

Fig. 4 illustrates a comparison between the present DDCS ( $e^+$ energy distribution ) and the experimental longitudinal $e^+$ distributions ( in arbitrary units [ 11 ] ) . The present DDCS are normalized ( scaled up by a factor of 1.5 ) to the experimental cross sections at the maxima of the distributions. As is revealed from the figures the present DDCS peak ( $e^+$ energy distribution ) shifts towards the higher value of the $E_{res}$ / 2 in contrast to the experimental shift [ 11 ] , which occurs at a lower value of $E_{res}$ / 2 . It may be mentioned in this context that the theoretical results of Starret et al [ 25 ] more or less agree qualitatively in this respect with the experiment [ 11 ] .

Finally, fig. 5 displays the DDCS summed over $\theta_1$, $\phi_1$ ( direct as well as exchange ) as a function of the ratio of the ejection velocity to the scattered velocity ( $R = \frac{v_e}{v_p}$ ) for forward emission of the electron ( i.e., $\theta_2 = 0^0$ ) , at different incident energies. Fig. 5 clearly demonstrates that the peak of the DDCS curves occur at a lower ratio of $\frac{v_e}{v_p}$ than unity ( i.e., $\frac{v_e}{v_p} < 1$ ) , particularly for lower incident energies ( e. g. , $E_i$ = 33 eV ) . However, the position of the DDCS peak shifts towards the higher ratio of $\frac{v_e}{v_p}$ with increasing incident energy and attains a value of almost unity ( $\frac{v_e}{v_p} \approx 1$ ) at $E_i$ = 100 eV ( vide fig. 5 ) . This shifting is in conformity with the experimental findings of Shah et al [ 12 ] for heavy particle impact. The occurrence of the DDCS peak value at a lower ratio of $\frac{v_e}{v_p}$ ( i.e., $\frac{v_e}{v_p} < 1$ ) could again be attributed to the post collisional interactions .

## 3. Conclusions

The salient features of the present ELC study are outlined below.

The angular distribution of the electron / positron ( TDCS ) exhibits a broad cusp like structure ( unlike the sharp cusp for heavy projectile ) at around half the residual energy ( $E_{res}/2$ ) for forward emission of the $e^-$ and the $e^+$ ( $\theta_1 = \theta_2 = 0^0$ ) indicating the occurrence of the ELC phenomena. The broadness of the cusp could be ascribed to the larger deflection of the light projectile Ps as compared to the forward scattering of the heavy projectile. The peak in the electron energy distributions ( DDCS ) occurs at below half the residual energy ( $E_{res}/2$ ) particularly at lower incident energies . The position of the DDCS peak ( summed over $\theta_1$, $\phi_1$ ) for forward electron emission shifts gradually towards the higher value of the ratio $\frac{v_e}{v_p}$ with increasing incident energy corroborating qualitatively the experimental observations for both heavy and light projectiles.

Finally, the proper judgment of a theoretical model needs a more rigorous absolute measurement for the lower level differential cross sections e.g., DDCS, TDCS for a wider incident energy range .

**Figure Captions:**

**Figure 1.** Triple differential cross sections ( TDCS ) in units of ( a. u. ) against the scattered positron angles ($\theta_1$) and for different values of the ejected electron angles ($\theta_2$). The incident positronium energy is 33 eV, scattered positron and ejected electron energies are 13 eV and 13.2 eV respectively. Dotted curve with open circle for $\theta_2 = 0^0$, solid curve for $\theta_2 = 30^0$, dashed curve for $\theta_2 = 60^0$, dotted line for $\theta_2 = 90^0$, dashed double dot curve for $\theta_2 = 120^0$ and dotted curve with full circle for $\theta_2 = 180^0$. Inset represents the same TDCS results due to Starrett etal [ 25 ].

**Figure 2.** TDCS ( in units of a.u ) against the scattered positron angle ( $\theta_1$ ) for different incident energies where the ejected electron angle ($\theta_2$) is fixed at $0^0$. Both the energies of the ejected electron and that of the scattered positron are equal. Solid curve for $E_i$ = 18 eV, $E_1 = E_2$ = 5.6 eV, dotted curve for $E_i$ = 50 eV, $E_1 = E_2$ = 21.6 eV. and dashed curve for $E_i$ = 200 eV, $E_1 = E_2$ = 96.6 eV.

**Figure 3. a.** Double differential cross sections ( DDCS ) summed over the $e^+$ scattering angles ( $\theta_1, \phi_1$ ) against the ejected ( scattered ) electron ( positron ) energy for electron emission angle $\theta_2 = 0^0$, at incident energy $E_i$ =18 eV. Lower abscissa represents the electron energy while the upper abscissa corresponds to the positron energy in eV. Solid curve represents the present direct result, dotted curve represents the present FBA results, dashed curve for exchange interaction in the frame work of Ochkur Rudge approximation [ 27, 28 ]. Inset exhibits the 3D DDCS ( both energy and angular distributions ) of Starrett et al [ 25 ].

**b.** Same DDCS but at higher incident energies $E_i = 500$ eV, 200 eV and 100 eV . Solid curves for direct results, dashed curves for exchange results in the frame work of Ochkur Rudge approximation [ 27, 28 ]. **c.** Same DDCS but for different values of the electron emission angles at incident energy $E_i = 33$ eV. Lines with solid circles for $\theta_2 = 0^0$. Solid lines for direct results, dashed lines for exchange results due to Ochkur Rudge approximation [ 27, 28 ] . Lines with open circles for $\theta_2 = 10^0$, lines with triangle for $\theta_2 = 45^0$. Same notations are used for direct and exchange results.

**Fig. 4.** DDCS summed over the e⁻ scattering angles ( $\theta_2, \phi_2$ ) against the scattered $e^+$ energy for $e^+$ emission angle $\theta_1 = 0^0$ at incident energies $E_i = $ 25 eV & 33 eV & the longitudinal energy spreads of $e^+$ due to Armitage etal [ 11 ] .

**Fig.5 .** The DDCS ( summed over the positron scattering angles ) is plotted against the ratio $\frac{v_e}{v_p}$ for two different incident energies. The electron emission angle is fixed at $0^0$. Line with open circles for $E_i = 100$ eV, line with solid circles for $E_i = 33$ eV, solid lines for direct result, dashed lines due to Ochkur Rudge approximation [ 27, 28 ].

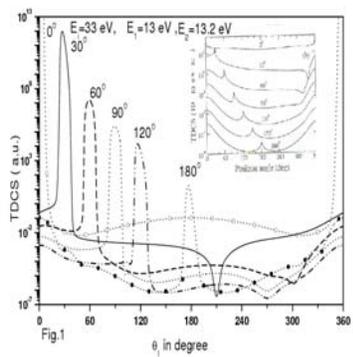

Fig.1

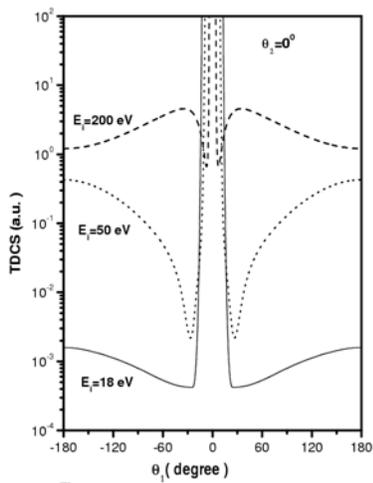

Fig 2.

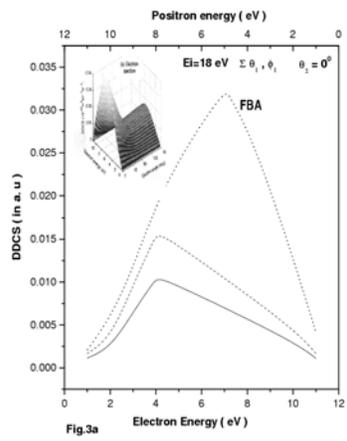

Fig.3a

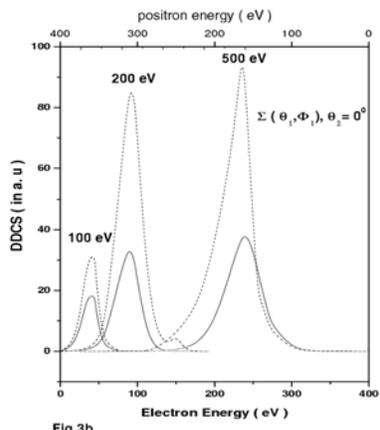

Fig.3b

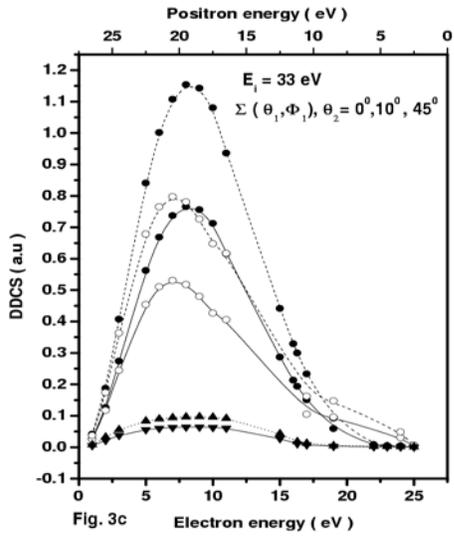

Fig. 3c

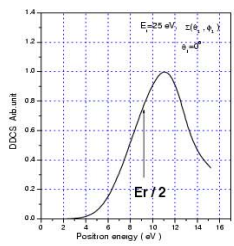 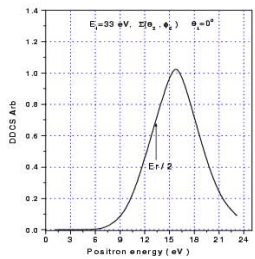
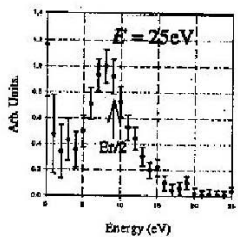 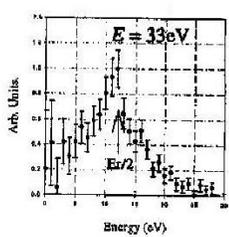

**Fig.4**

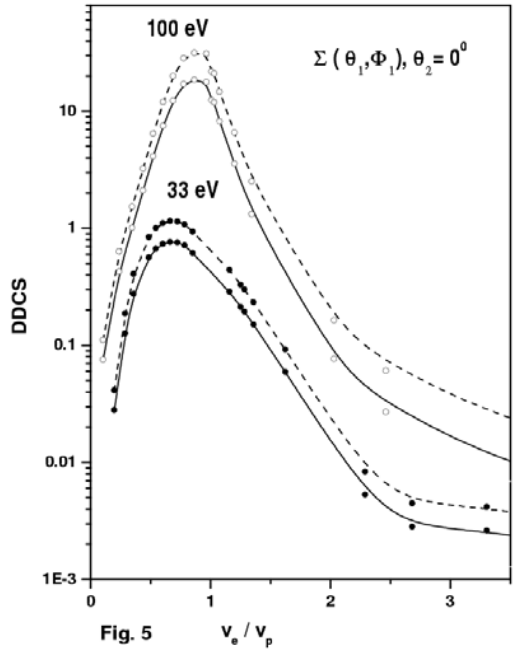

Fig. 5